\documentclass[a4paper,10pt, onecolumn]{nveart}
\usepackage{hyperref} 
\usepackage{amsfonts} 
\usepackage{amssymb} 
\usepackage{amsthm} 
\usepackage{makeidx} 
\usepackage{enumerate} 
\usepackage{graphicx} 
\usepackage{amsmath} 
\usepackage{fancyhdr}
\usepackage{multicol}
\usepackage{fullpage}
\usepackage{color}
\usepackage{units}
\usepackage{lineno}

\newcommand{\ie}{\textit{i.e. }}

\newcommand{\be}{\begin{equation}}
\newcommand{\ee}{\end{equation}}
\newcommand{\best}{\begin{equation}\rm} 
\newcommand{\eest}{\end{equation}} 
\newcommand{\benum}{\begin{enumerate}}
\newcommand{\eenum}{\end{enumerate}}
\newcommand{\bitm}{\begin{itemize}}
\newcommand{\eitm}{\end{itemize}~\\}

\newcommand{\gr}[1]{\textbf{#1}} 
\newcommand{\rf}[1]{(\ref{#1})} 

\newcommand{\nli}{\\\indent} 

\newcommand{\remarques}{~\nli\begin{small}\textsc{remarques :}\\\benum[(i)]} 
\newcommand{\ermarques}{\eenum\end{small}~\\\indent}

\newcommand{\rappels}{~\nli\begin{small}\textsc{rappels :}\\\benum[(i)]} 
\newcommand{\erappels}{\eenum\end{small}~\\\indent}

\renewcommand{\thefootnote}{\fnsymbol{footnote}}

\newtheorem{Demonstration}{\indent DŽmonstration}

\journal{Astroparticle Physics}

\begin{document}

\begin{frontmatter}
\title{Bayesian Approach for Counting Experiment Statistics applied to a Neutrino Point Source Analysis}
\author{D. Bose, L. Brayeur, M. Casier, K. D. de Vries, G. Golup and N. van Eijndhoven}
\address{Vrije Universiteit Brussel, Dienst ELEM\\Pleinlaan 2, B-1050 Brussels, Belgium}

\begin{abstract}
In this paper we present a model independent analysis method following Bayesian statistics to analyse data from a generic counting experiment and apply it to the search for neutrinos from point sources. We discuss a test statistic defined following a Bayesian framework that will be used in the search for a signal. In case no signal is found, we derive an upper limit without the introduction of approximations. The Bayesian approach allows us to obtain the full probability density function for both the background and the signal rate. As such, we have direct access to any signal upper limit. The upper limit derivation directly compares with a frequentist approach and is robust in the case of low-counting observations. Furthermore, it allows also to account for previous upper limits obtained by other analyses via the concept of prior information without the need of the \textit{ad hoc} application of trial factors.
To investigate the validity of the presented Bayesian approach, we have applied this method to the public IceCube 40-string configuration data for 10 nearby blazars and we have obtained a flux upper limit, which is in agreement with the upper limits determined via a frequentist approach. Furthermore, the upper limit obtained compares well with the previously published result of IceCube, using the same data set.
\end{abstract}


\end{frontmatter}

\section{Introduction}
Bayesian approaches are gaining more and more popularity in scientific analyses \cite{jaynes,Feigelson,Chiara,D'ag,cousins}. In this paper we discuss a formalism following Bayesian inference \cite{loredo,greg} to analyse signals in a generic counting experiment and for illustration we apply it to a point source analysis using data from a neutrino telescope.

A frequentist approach is based on the long run relative frequency of occurrences in identical repeats of an experiment. Consequently, this can only provide the probability for a certain outcome under the assumption of a specific hypothesis. On the other hand, the Bayesian approach allows us to directly compute the probability of any particular hypothesis or parameter value based on observations. One of the great strengths of Bayesian inference is the ability to incorporate relevant prior information in the analysis. This provides a mechanism for a statistical learning process that automatically takes previous results into account.

Concerning a signal detection we will analyse the performance of the presented Bayesian approach and we will compare it to methods frequently found in literature. One of the standard test statistics is the frequentist method developed by Li and Ma \cite{lima}. While this method is based on the detection of an excess of number of events in a given angular window, the presented Bayesian method is sensitive to the distribution of the angular distance between the arrival direction of the observed events and the source position.

In case no signal is observed, upper limits for the signal flux from the investigated point sources are determined. In the light of a flux upper limit determination we will discuss a method following the Bayesian framework and a comparison will be made to the standard frequentist methods developed by Feldman and Cousins \cite{feldman} and Rolke \textit{et al.} \cite{rolke}. The Feldman-Cousins method introduces the likelihood ratio as an ordering principle, when determining the acceptance interval from which one derives the confidence interval. This construction resolves the issue of empty confidence intervals when the interval is entirely in the non-physical region. The Feldman-Cousins construction also resolves the ``flip-flop'' problem, which can lead to under coverage. This arises when one wants to report an upper limit if the evidence for a signal is below a certain threshold, but a central confidence interval otherwise. The procedure outlined by Rolke \textit{et al.} can deal with several nuisance parameters which include uncertainties (for example background expectations or signal efficiencies) by means of the Profile Likelihood method. However, this method has the disadvantage that large samples are assumed and thus applying it to cases with low counting statistics might lead to under coverage.

The Bayesian method presented here will be applied to the study of Active Galactic Nuclei (AGN), which are, together with Gamma Ray Bursts (GRBs), among the leading candidates for the sources of ultra-high energy cosmic rays (UHECRs), as outlined in recent reviews such as \cite{selvon,olinto}. On time scales of order of minutes or even seconds GRBs are transient phenomena, whereas AGN in general can be regarded as steady sources over time periods even exceeding several years. In case hadronic acceleration takes place in these objects, also an accompanying high-energy neutrino flux is expected due to the decay of secondary particles produced in the interactions of accelerated hadrons with the ambient photon field or matter \cite{waxman}.

Kilometer-scale neutrino detectors have the sensitivity to measure the predicted neutrino flux. The IceCube Neutrino Telescope \cite{genIceCube}, located at the geographic South Pole, was completed at the end of 2010 and has been taking data since 2006. In the Northern Hemisphere a kilometer-scale detector called KM3Net is being proposed for construction \cite{km3net}. There are smaller telescopes, one located in the Mediterranean Sea, named ANTARES which has been collecting data since 2008 \cite{antares} and another one in Lake Baikal, Russia, which has been operating its NT$200+$ configuration since 2005 \cite{baikal}. These telescopes detect high energy neutrinos ($E>10$ GeV) by observing Cherenkov radiation in ice or water from secondary particles produced in neutrino interactions.

In the following we introduce the basis of the Bayesian formalism and present the test statistic that we will use to assess consistency of the data with the null hypothesis, being an isotropic distribution of the events. In Section 4, we describe the procedure to determine upper limits in case no significant signal is observed. Subsequently, in Section 5, we apply the method to IceCube public data \cite{data} investigating the neutrino flux from the ten closest blazars (a subclass of AGN). Finally, a summary of our results and conclusions are provided in Section 6.

\section{Bayesian Formalism}
We denote by $p(A,B|C)$ the probability for hypothesis $A$ and $B$ to be true under the condition that $C$ is true. The product rule \cite{jaynes}, $p(A,B|C)=p(A|C)p(B|A,C)=p(B|C)p(A|B,C)$, directly yields Bayes' theorem \cite{hayes}:
\be
p(A|B,C)=\frac{p(A|C)p(B|A,C)}{p(B|C)}.
\ee

Bayes' theorem is extremely powerful in hypothesis testing. Consider a hypothesis $H$, some observed data $D$
and prior information $I$. Bayes' theorem can then be rewritten as:
 \begin{equation} \label{b1}
   {
  p(H|D,I)=\frac{p(H|I) p(D|H,I)}{p(D|I)},
   }
 \end{equation}

where\\

 $\begin{array}{lll}
  \ \ \ \ p(H|D,I) &\equiv&\textrm{ Posterior probability of hypothesis }H. \\
  \ \ \ \ p(H|I) &\equiv&\textrm{ Prior probability of hypothesis }H.\\
  \ \ \ \ p(D|H,I) &\equiv&\textrm{ Likelihood function, }\mathcal{L}(H). \\
  \ \ \ \ p(D|I) &\equiv&\textrm{ Normalization factor.}\\
\end{array}$
\\

From Eq. \rf{b1} it is seen that the Bayesian formalism automatically provides a learning process. The first step is to encode our state of knowledge before analyzing the data into a prior probability $p(H|I)$. This is then converted into a posterior probability $p(H|D,I)$ when a new data set is analysed.

We will apply this approach to the analysis of astrophysical point sources. This method will enable us to obtain the full Probability Density Function (PDF) for the source rate (i.e. the number of signal events per unit of time arriving from the source) from which we can derive the corresponding flux or any upper limit.

\section{Assessment of Significance}

As outlined above, Bayesian inference allows us to make statements about the probability of various hypotheses in the light of obtained data. Following the developments described in \cite{NvE}, we quantify the degree to which data support a certain hypothesis and as such make an assessment of the significance. To quantify our degree of belief in a certain hypothesis $H$, one can use the so called evidence \cite{jaynes}
\be
e(H|D,I)=10\log_{10}\Bigg[\frac{p(H|D,I)}{p(\overline{H}|D,I)}\Bigg],
\label{BH}
\ee
where $\overline{H}$ indicates hypothesis $H$ to be false. Due to the $\log_{10}$, this evidence is in a decibel scale. One can calculate \cite{NvE} that the evidence $e(\overline{H}|D,I)$ for any alternative to hypothesis $H$ based on the data $D$ and prior information $I$ is constrained by the observable
\be
\psi \equiv -10\log_{10}p(D| H,I)
\ee
Thus, $\psi$ provides a reference to quantify our degree of belief in $H$.

Let us now consider an experiment where the probabilities $p_{k}$ corresponding to the various outcomes $A_{k}$ on successive trials are independent and stationary. Such experiments belong to the so called Bernoulli class $B_{m}$ \cite{jaynes}. The probability $p(n_{1}....n_{m} | B_{m},I)$ of observing $n_{k}$ occurrences of each outcome $A_{k}$ after $n$ trials is therefore given by the multinomial distribution:
\begin{equation} \label{b6}
 {
  p(D | B_{m},I)=p(n_{1}....n_{m} | B_{m},I)=\frac{n!}{n_{1}!....n_{m}!} p_{1}^{n_{1}}....p_{m}^{n_{m}}.
 }
\end{equation}
In terms of the observable $\psi$ we then obtain \cite{NvE} for each $H\in B_m$:
\begin{equation} \label{b7}
 {
  \psi_{B_m} = -10 \left [ \log_{10} n! + \sum_{k=1}^{m} \left ( n_{k} \log_{10} p_{k} - \log_{10} n_{k}! \right ) \right ],
 }
\end{equation}
which is an exact expression.

In case the data are represented in histogram form, the above can be applied if $n$ is the total number of entries, $m$ represents the number of bins, $n_{k}$ is the number of entries in bin $k$ and $p_{k}$ is the probability for an entry to fall in bin $k$. Once the various $p_k$ are known, the $\psi$-value corresponding to a certain observed distribution can easily be obtained using Eq. \rf{b7}.

In our investigation of neutrino point sources, the total number of events $n$ will populate a histogram according to the angular difference $\alpha$ of each neutrino arrival direction with respect to the source location in the sky. The $p_k$ values are determined assuming an isotropic background and taking into account the solid angle effect. By generating randomly the same total number of events\footnote{The number $n$ has to be fixed due to the first term in the definition of $\psi$, Eq. \rf{b7}. This implies that $\psi$ is sensitive to the shape of the distribution of events within the distance $\alpha$ but not to the total number of events, \ie if there is an excess of signal events that follows an isotropic distribution, the $\psi$ observable will not enable us to distinguish it from background.}following an isotropic distribution, we obtain the distribution of $\psi$ for an isotropic background. Comparing the value of $\psi$ for the data with the obtained background $\psi$-distribution, we determine the P-value or significance of our measurement. A source detection is claimed if the consistency of the data with the null hypothesis has a P-value smaller than $5.7 \times 10^{-7}$, corresponding to a $5\sigma$ effect in case of a positive single sided Gaussian distribution.

It is important to note that as $\psi$ is sensitive to the distribution of the data in the histogram, it is binning-dependent. To avoid this binning effect in the final P-value, the bin size is chosen such that there is only one or zero event per bin. In this special, quasi-unbinned approach, Eq. \rf{b7} would simplify in the following expression
\be
\psi_{B_m}=-10 \left[ \log_{10} n! + \sum_{k=1}^{n} \log_{10} p_k \right],
\label{b7reduced}
\ee
where the sum is running over the total number of events $n$.

The $p_k$ probabilities are then the crucial factor that will differentiate between background events following an isotropic distribution and signal events from a source. The signal events will be located at angular distances around the source position following a Gaussian distribution with its standard deviation given by the experimental angular resolution convoluted with the solid angle effect. 

On the other hand, the test statistic developed by Li and Ma \cite{lima} is based on the absolute value of the number of events in the on-source ($N_\textrm{on}$) and off-source ($N_\textrm{off}$) angular windows. Consequently, both test statistics behave differently as a function of the considered angular window. As we will show in Section 5, the Li-Ma test statistic performs better than $\psi$ for small angular windows while $\psi$ is more sensitive at larger windows. Note that in Section 5 the complete expression of Eq. \rf{b7} is used instead of the simplified Eq. \rf{b7reduced} since for our background distributions it occurs that $n_k>1$ in a small fraction of cases, which we control to be at maximum 5\%.

In case the observation does not lead to a significant detection, we determine an upper limit on the signal strength. To obtain such an upper limit we will use an exact analytical expression following a Bayesian approach using a uniform prior. The details of this procedure are provided in the following section.

\section{Bayesian upper limit determination}

To obtain an upper limit for a possible source flux, we first have to determine the upper limit for the source rate. Using Bayesian inference we obtain the full posterior PDF for the source rate and from that we can derive the corresponding flux PDF, via the concept of effective area, as explained hereafter.

In any experiment where events are detected at a known rate and independently of the time since the last event, the PDF for the number of observed events is described by a Poisson distribution. Our approach to obtain an upper limit follows the one outlined in \cite{loredo,greg}, except that we apply the exact analytical expression without any approximation.

The data consist of on-source and off-source measurements, where the off-source data consist only of background events and the on-source data are a mix of background and source (\ie signal) events. We start with the off-source analysis and subsequently use the obtained information as background prior information in the on-source analysis. This implies that the current study is completely data driven and as such is a model independent search for a possible source signal.

\subsection{Off-source measurements}

Consider the case that in an off-source measurement $N_\textrm{off}$ background events have been recorded over a time interval $T_\textrm{off}$ with a constant background rate $b$. Using Eq.($\ref{b1}$) we obtain the posterior background PDF by:
\begin{equation} \label{b8}
 {
 p(b| N_\textrm{off},I)=\frac{p(b| I) p(N_\textrm{off}| b,I)}{p(N_\textrm{off}|I)}.
 }
\end{equation}

In the above equation the likelihood function $ p(N_\textrm{off}| b,I)$  is given by the Poisson distribution corresponding to the measurement of $N_\textrm{off}$ background events over a time span $T_\textrm{off}$ at a constant rate $b$:
\begin{equation} \label{b9}
 {
 p(N_\textrm{off}| b,I) = \frac{(bT_\textrm{off})^{N_\textrm{off}}e^{-bT_\textrm{off}}}{N_\textrm{off}!}.
 }
\end{equation}

Since the integrated PDF amounts to 1 (\ie $ \int_{b_\textrm{min}}^{b_\textrm{max}} p(b| N_\textrm{off},I)\,\textrm{d}b= 1$), the normalisation factor $p(N_\textrm{off}|I)$ appearing in Eq. \rf{b8} is given by:
\begin{equation} \label{b11}
 {
 p(N_\textrm{off}|I) = \int_{b_\textrm{min}}^{b_\textrm{max}} p(b| I) p(N_\textrm{off}| b,I) \textrm{d}b.
 }
\end{equation}

As mentioned before, $p(b| I)$ is the prior PDF for the background rate. For our analysis we use a uniform prior \cite{greg}, which is given by
\begin{equation} \label{b12}
 {
 p(b| I) = \frac{1}{b_\textrm{max}-b_\textrm{min}}.
 }
\end{equation}

The uniform prior attributes the same probability to every value of the background rate within the indicated range, reflecting that we do not favour a particular value of the actual background rate. This prior also has the advantage that the derived upper limits are directly comparable to classical frequentist upper limits \cite{cousins}.

To cover the full range of possible background rates, the minimum value of the rate $b$ is taken to be zero and Eq. ($\ref{b12}$) can be written as
 \be
 p(b| I) = \frac{1}{b_\textrm{max}}.
 \label{b12bis}
 \ee

 Using this expression for $p(b| I)$ together with Eqs. \rf{b9} and \rf{b11} we obtain an analytical expression for the normalisation factor
\begin{equation} \label{b13}
 {
 p(N_\textrm{off}|I) = \int_{0}^{b_\textrm{max}} \frac{1}{b_\textrm{max}} \frac{(bT_\textrm{off})^{N_\textrm{off}}e^{-bT_\textrm{off}}}{N_\textrm{off}!} \textrm{d}b.
 }
\end{equation}

 Solving the above equation (for details see Appendix A1) we get:
\begin{equation} \label{b14}
 {
 p(N_\textrm{off}|I) = \frac{1}{b_\textrm{max}} \frac{\gamma(N_\textrm{off}+1, b_\textrm{max}T_\textrm{off})}{N_\textrm{off}!\,T_\textrm{off}},
 }
\end{equation}
where $\gamma(a,x)$ is the Incomplete Gamma Function. Substitution of Eqs. ($\ref{b9}$), \rf{b12bis} and ($\ref{b14}$) in Eq. ($\ref{b8}$) yields
\begin{equation} \label{b15}
 {
 p(b| N_\textrm{off},I)= \frac{T_\textrm{off}(bT_\textrm{off})^{N_\textrm{off}}e^{-bT_\textrm{off}}}{\gamma(N_\textrm{off}+1, b_\textrm{max}T_\textrm{off})},
 }
\end{equation}
which represents the posterior background rate PDF. In \cite{loredo,greg} the Incomplete Gamma Function is then approximated to $\gamma(N_{\rm off}+1, b_{\rm max} T_{\rm off}) \approx \Gamma (N_{\rm off}+1) = N_{\rm off}!$. This approximation is valid for $T_{\rm off}\,b_{\rm max} \gg N_{\rm off}$ and cannot be applied in general when the number of events and time window are small (see \cite{NvE} for an example). We therefore work with the complete analytical expression.

\subsection{On-source measurements}

Consider the case that in an on-source measurement $N_\textrm{on}$ events, consisting of signal and background, have been recorded over a time interval $T_\textrm{on}$ with a constant signal rate $s$ and background rate $b$. Following Eq. \rf{b1} the joint probability of source and background is given by:
\begin{equation} \label{b16}
 {
 p(s,b | N_\textrm{on},I) = \frac{p(s,b | I) p(N_\textrm{on} | s,b,I)}{p(N_\textrm{on} | I)}.
 }
\end{equation}

 Using the product rule \cite{jaynes} we can write the above equation as,
\begin{equation} \label{b17}
 {
 p(s,b | N_\textrm{on},I) = \frac{p(b | I) p(s | b,I)p(N_\textrm{on} | s,b,I)}{p(N_\textrm{on} | I)},
 }
\end{equation}

 where $p(b | I)$ is the prior probability for the background rate, which is in our case the posterior background PDF obtained from the off-source measurement reflected in Eq. ($\ref{b15}$). The likelihood function $p(N_\textrm{on} | s,b,I)$ is the Poisson distribution for the combined signal and background rate $(s+b)$. The normalisation constant  $p(N_\textrm{on} | I)$ is obtained, as outlined in the previous subsection, by integrating the numerator of Eq. ($\ref{b17}$).

It is important to note that since the source rate $s$ and the background rate $b$ are independent, we can write $p(s | b,I)=p(s |I)$. Like for the background case discussed before, we use a uniform prior for $p(s | I)$, \ie
\begin{equation} \label{b18}
 {
 p(s | I) = \frac{1}{s_\textrm{max}}=p(s | b,I).
 }
\end{equation}
As mentioned before, the uniform prior attributes the same probability to every value of the signal rate within the indicated range, reflecting that we do not favour a particular value of the actual signal rate and also allows us to directly compare the derived upper limits with the classical frequentist results \cite{cousins}.

By substituting in Eq. \rf{b17} the expressions of Eq. \rf{b18} for $p(s|b,I)$, \rf{b15} for $p(b|I)$ and \rf{b9} for a total rate $(s+b)$ we obtain the joint PDF for the source and background rates:
\be
p(s,b|N_\textrm{on},I)=\frac{(bT_\textrm{off})^{N_\textrm{off}}\,e^{-bT_\textrm{off}} \cdot(b+s)^{N_\textrm{on}}\,T_{\rm on}^{N_{\rm on}}\,e^{-(b+s)T_\textrm{on}}}{\int_0^{b_\textrm{ max}}\int_0^{s_\textrm{ max}}(bT_\textrm{off})^{N_\textrm{off}}\,e^{-bT_\textrm{off}}\cdot(b+s)^{N_\textrm{on}}\,e^{-(b+s)T_\textrm{on}}\,\textrm{d}b\,\textrm{d}s}.
\label{pdfsb}
\ee

However, we are interested in the posterior PDF for the source rate alone, independent of the background. The Bayesian formalism allows us to obtain this posterior PDF by marginalisation \cite{jaynes} of the joint PDF, Eq. \rf{pdfsb}, with respect to the background \ie
\begin{equation} \label{b19}
 {
 p(s | N_\textrm{on},I) = \int_{0}^{b_\textrm{max}} p(s,b | N_\textrm{on},I) \textrm{d}b.
 }
\end{equation}

Solving the above integral we obtain an exact expression for the source rate posterior PDF (for details see Appendix A2):
\begin{equation} \label{b20}
 {
 p(s | N_\textrm{on},I) = \frac{e^{-sT_\textrm{on}} \sum_{i=0}^{N_\textrm{on}} \frac{s^{i}(T_\textrm{on}+T_\textrm{off})^{i}\gamma(N-i+1,u_\textrm{max})}{i!(N_\textrm{on}-i)!}} {\sum_{j=0}^{N_\textrm{on}} \frac{(T_\textrm{on}+T_\textrm{off})^{j}\gamma(N-j+1,u_\textrm{max})\gamma(j+1,s_\textrm{max}T_\textrm{on})}{j!(N_\textrm{on}-j)!T_\textrm{on}^{j+1}}}\,,
 }
\end{equation}
where $ u_\textrm{max} \equiv b_\textrm{max}(T_\textrm{on}+T_\textrm{off})$ and $ N \equiv N_\textrm{on}+N_\textrm{off} $. As explained before we use this exact analytical expression without the approximation used in \cite{loredo,greg}.

In case no significant source signal is observed, Eq. \rf{b20} allows us to derive any upper limit for the source rate. As an example, the 90\% source rate upper limit $s_\textrm{u.l.}$ is given by:
\be
\int_0^{s_\textrm{u.l.}}p(s | N_\textrm{on},I) \textrm{d}s=0.9
\label{upper limit}
\ee

\section{Application of the method to a neutrino point source analysis}

The approach presented in the current paper is not exploiting a potential source variability when it comes to providing upper limits to a possible source rate, as outlined in Section 6. Consequently the analysis presented here is tailored for sources with a steady rate within the detection time. Our primary goal is the analysis of AGN which, at the time scale considered here, may be regarded to be steady sources of high-energy neutrinos if hadronic acceleration takes place in these sites. By selecting a small region around every well known source location, several sky patches are defined from which data were collected to search for a possible deviation from the background ``noise".

To validate the analysis procedure described in this report, we use the public data \cite{data} of the muon neutrino candidate events recorded by the IceCube Neutrino Observatory \cite{genIceCube,sensitivity} in its 40 string configuration (IC40), that collected data during the season 2008-2009. Our analysis is performed on ten nearby blazars (a special class of AGN with one of the jets pointing in the direction of the Earth) following the approach described in the previous sections. The blazars were selected from the online ``Roma BZCAT Multi-frequency Catalogue of Blazars" \cite{agnlist} and are listed in Table \ref{blazars}. These blazars are chosen to be nearby, \ie with a small redshift, and in such a way that their respective angular windows are not overlapping. We limit ourselves to sources in the Northern hemisphere to reduce the atmospheric muon background for the IceCube measurements.

\begin{table}[h!]
\centering
   \begin{tabular}{|c|c|c|c|c|}
  \hline
  Blazar Name&Right Ascension&Declination&Redshift&Distance [Mpc]\\
  \hline
  \hline
BZUJ1148+5924&177.20983&59.41567&0.011&46.2\\
BZUJ0048+3157&12.19642&31.95697&0.015&63\\
BZUJ0319+4130&49.95067&41.51169&0.018&75.6\\
BZUJ0709+5010&107.39246&50.18225&0.02&84\\
BZUJ0153+7115&28.35771&71.25181&0.022&92.4\\
BZUJ1719+4858&259.81025&48.98042&0.024&100.8\\
BZUJ1632+8232&248.13321&82.53789&0.025&105\\
BZUJ1755+6236&268.95183&62.61225&0.027&113.4\\
BZBJ1104+3812&166.11379&38.20883&0.030&126\\
BZBJ1653+3945&253.46758&39.76017&0.033&139.3\\
  \hline
\end{tabular}
\caption{\label{blazars}Nearby blazars \cite{agnlist} used in the current analysis.}
\end{table}

\subsection{Assessment of significance}

As outlined in \cite{NvE}, we stack the recorded events within a given angular window centered on each of these ten blazars according to their angular distance $\alpha$ from the actual blazar position. As mentioned in the previous sections, we need an expression for the various probabilities $p_k$ of Eq. \rf{b7} to derive the $\psi$-value of the signal. In our case, the background is isotropic and consequently the probabilities have to be consistent with the solid angle effect within the selected cone, \ie:
\be
p_k=\frac{1}{1-\cos(\alpha_{\rm max})}[\cos(w\cdot k)-\cos(w\cdot(k+1))]\,,
\label{pk}
\ee
where $w$ is the width of each bin in our stacked histogram and $\alpha_{\rm max}$ is the size of the angular window.\\

\begin{figure}[h!]
\centering
\includegraphics[scale=0.7]{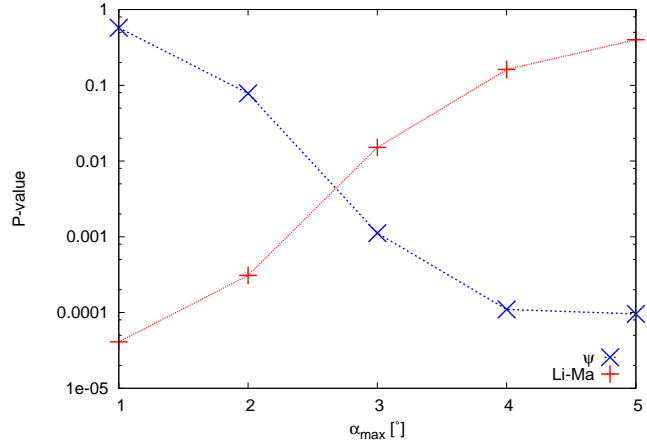}
\caption{\label{psilima} Mean P-values obtained with the simulated signals as a function of the angular window ($\alpha_{\rm max}$) for the $\psi$ and Li-Ma methods (assuming an experimental angular resolution of 1 degree).}
\end{figure}

To determine the angular window size for which our test-statistic is most sensitive, we have generated 266 (being the number of events of the actual observation as outlined hereafter) isotropically distributed events plus a signal of 20 events. These 20 events were generated such that their angular distance $\alpha$ from the source position follows a Gaussian distribution (with standard deviation of 1 degree) convoluted with the solid angle effect. The chosen standard deviation of the Gaussian distribution is the angular uncertainty of the IceCube track reconstruction \cite{sensitivity,moon}. As we mentioned before, our test statistic is sensitive to the distribution of events as a function of the distance to the source. To avoid the dependence on the position generation in our determination of the most optimal angular window size, we have repeated the procedure 10 times for each value of $\alpha$ for the signal events and present the mean P-values in Fig. \ref{psilima}.

From Fig. \ref{psilima} it is seen that our $\psi$ test statistic does not perform well on small angular windows. This is due to the fact that the difference of the individual probabilities per bin ($p_k$) for small angular windows is not large enough to distinguish a source-like distribution from an isotropic background (\ie a larger angular window is needed to see the ``shape" of the excess). For an angular window of 4 degrees, we see that the sensitivity obtained with $\psi$ becomes optimal.

We also include in Fig. \ref{psilima} a comparison to the P-value obtained with the standard Li-Ma method. As mentioned in Section 3, Li-Ma is a test statistic based on the total number of events in the on-source ($N_\textrm{on}$) and off-source ($N_\textrm{off}$) angular windows. Li-Ma performs better than $\psi$ for small angular windows (a large $N_{\rm on}/N_{\rm off}$ ratio). When comparing the smallest P-values of each test statistic, we see that both are similar. 

\begin{figure}[!htb]
\centering
\includegraphics[scale=0.7]{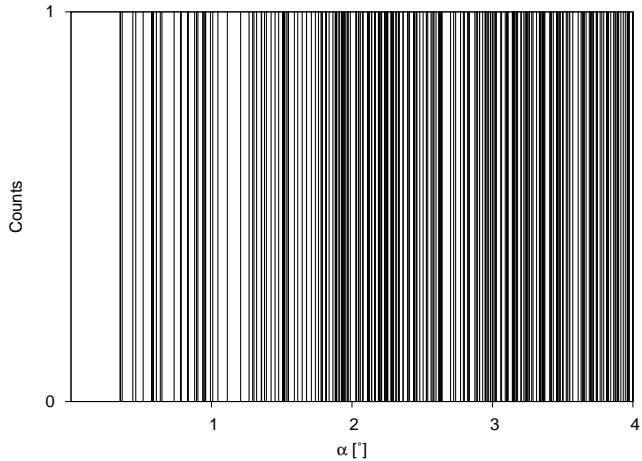}
\caption{The stacked distribution of events within a 4$^\circ$ cone of all our 10 Blazars of Table \ref{blazars}, with $\alpha$ the angle between the corresponding blazar location and the reconstructed arrival direction.}
\label{stackedblazar}
\end{figure}

Fixing $\alpha_{\rm max}$ to 4 degrees, we use the IceCube public data and we obtain the stacked distribution of events presented in Fig. \ref{stackedblazar}. These stacked data comprise 266 events recorded over a time period of 375.5 days \cite{diffuse}. Using Eqs. \rf{b7} and \rf{pk} for a number of entries $N_{\rm on}= 266$, the data represented in Fig. \ref{stackedblazar} yield $\psi_{observed} = 10621\,$dB.

\begin{figure}[!htb]
\centering
\includegraphics[scale=0.7]{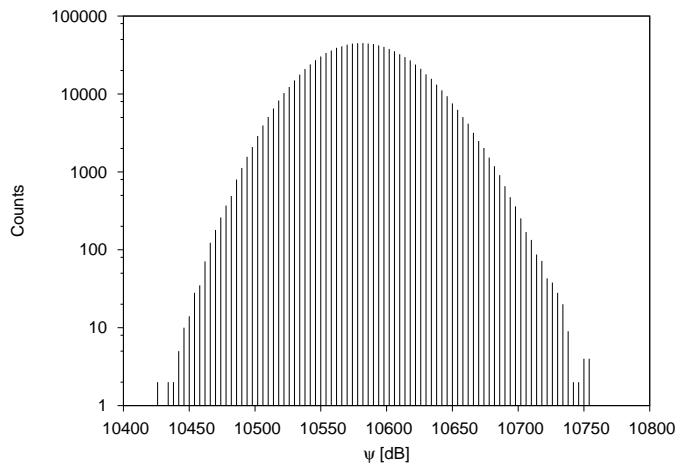}
\caption{\label{psidistribution}The distribution of $\psi$-values for $10^6$ generated isotropic background events.}
\end{figure}

As explained in Section 3, the $\psi$ distribution in the case of an isotropic background is obtained by randomly generating $10^6$ times the same number of events as in the on-source region.  The distribution is presented in Fig. \ref{psidistribution}. Comparison of the actual observation $\psi_{observed}$ with the background distribution $\psi_{bkg}$ gives a $P$-value of $0.15$. Consequently, we will proceed to give an upper limit on the signal strength. \\

\subsection{Upper limit determination using Uniform priors for Source and Background}

\subsubsection{Determination of the background rate}

\begin{figure}[!h]
\centering
\includegraphics[scale=0.7]{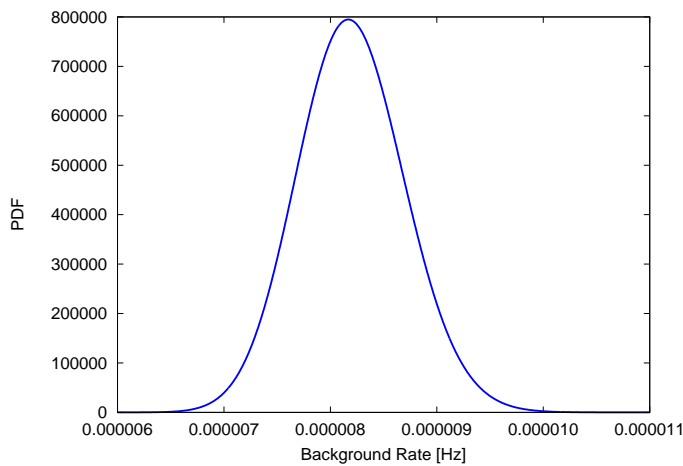}
\caption{\label{bckgrdblazar}Off-source rate PDF using a uniform prior. For the analysed data example: $N_{\rm off}=265$, $T_{\rm off}=375.5$ days and $b_\mathrm{max}=0.8$ mHz.}
\end{figure}

To determine the number of events in the off-source region ($N_{\rm off}$) we perform measurements in 4$^\circ$ regions of the sky, shifted from the various blazars positions only in right ascension, keeping the declination constant due to the declination dependence of the IceCube detection efficiency. The specific IC40 configuration of IceCube is also right ascension dependent, so we make shifts of 180$^\circ$ in right ascension to eliminate the right ascension dependence. The IC40 sample has been taken over a detector live time period of 375.5 days, so that both the exposures for on-source, $T_{\rm on}$, and off-source, $T_{\rm off}$, amount to 375.5 days. The stacked off-source measurements yield a total of 265 events. The posterior background rate PDF is obtained by substitution of the previously mentioned values of $T_\mathrm{off}$ and $N_\mathrm{off}$ in Eq. \rf{b15} and by using a sufficiently large value $b_\mathrm{max}=\frac{N_\mathrm{off}}{T_\mathrm{off}}\cdot 100 =$0.8 mHz. The resulting background rate PDF is shown in Fig. \ref{bckgrdblazar}.\\

\subsubsection{Determination of the source rate}

\begin{figure}[!h]
\centering
\includegraphics[scale=0.7]{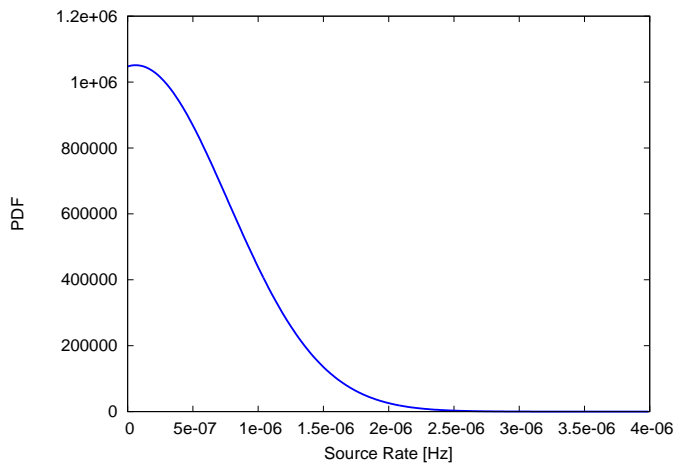}
\caption{\label{sourceblazar}Source rate PDF using a Uniform prior. For the analysed data example: $N_{\rm off}=265$, $N_{\rm on}=266$, $T_{\rm on}=T_{\rm off}=375.5$ days, $b_\mathrm{max}=0.8$ mHz and $s_\mathrm{max}=1$ Hz.}
\end{figure}

The posterior source rate PDF is obtained by inserting the previously mentioned values of $T_\mathrm{on}$, $T_\mathrm{off}$, $N_\mathrm{on}$ and $N_\mathrm{off}$ in Eq. \rf{b20} and by using a sufficiently large value $s_\mathrm{max}= 1$ Hz. The resulting source rate PDF is shown in Fig. \ref{sourceblazar}.

Using the PDF shown in Fig. \ref{sourceblazar} and applying Eq. \rf{upper limit}, we obtain the 90\% upper limit for the source rate:
\be
s_\mathrm{u.l.}=1.2\times 10^{-6}\,\mathrm{Hz}. \nonumber
\label{upperlimit}
\ee

To compare this Bayesian method with the frequentist approach, we have also determined the 90\% upper limit for the source rate using the Feldman-Cousins \cite{feldman} and Rolke \textit{et al.} \cite{rolke} methods. The values obtained are the following:

\begin{eqnarray}
&s_\mathrm{u.l.\,FC}&= 9.0 \times 10^{-7}\,\mathrm{Hz} \nonumber \\
&s_\mathrm{u.l.\,R}&= 1.2 \times 10^{-6}\,\mathrm{Hz}.\nonumber.
\end{eqnarray}

\begin{figure}[!h]
\centering
\includegraphics[scale=0.7]{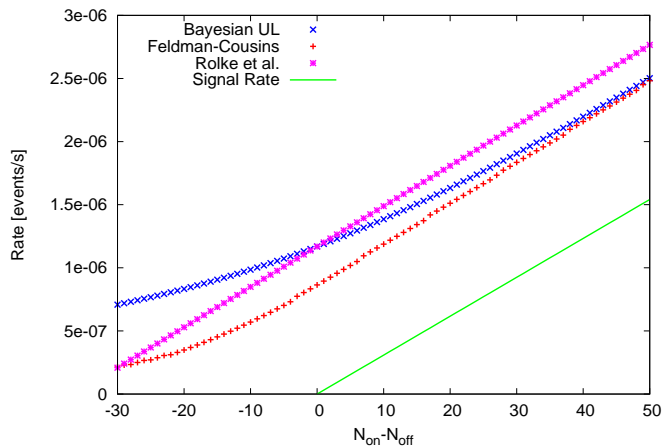}
\caption{\label{ULcomp}Comparison of the rate upper limits obtained with the Bayesian method and the frequentist Feldman-Cousins and Rolke \textit{et al} methods as a function of the difference of events between the on-source and off-source regions.}
\end{figure}

We see that the Bayesian approach is equal to the Rolke \textit{et al.} method and is more conservative than the Feldman-Cousins method. To further test the upper limit calculation, we have generated source signals (or under-fluctuations) by increasing (or decreasing) the number of events in the on-source region, while keeping the same number of background events as in the data. In Fig. \ref{ULcomp} we plot the rate upper limits obtained with the Bayesian and frequentist methods as a function of the difference of the number of events between the on-source and off-source regions ($N_{\rm on} - N_{\rm off}$). We also show the actual rate for the case of a positive difference of $N_{\rm on} - N_{\rm off}$. The Bayesian upper limits are similar to the Rolke \textit{et al.} limits for a small difference of $N_{\rm on} - N_{\rm off}$ but the former is more restrictive when this difference increases. When comparing to the Feldman-Cousins results, the Bayesian limits are more conservative for low $N_{\rm on} - N_{\rm off}$ and tend to the Feldman-Cousins limits as this difference grows, as expected from the fact that we use an uniform prior.

\begin{figure}[!h]
\centering
\includegraphics[scale=0.7]{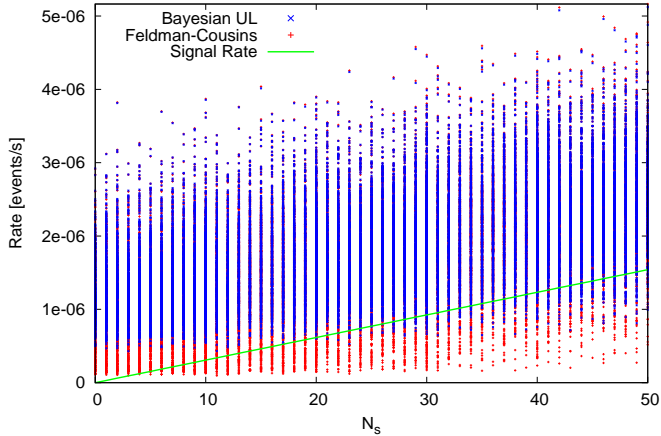}
\caption{\label{UpLimmore}Comparison of the rate upper limits obtained with the Bayesian method and the frequentist Feldman-Cousins method as a function of the number of source events.}
\end{figure}

The decrease of the slope when the difference of the number of events is negative shows that the Bayesian method is better protected against under-fluctuations. This effect is shown in Fig. \ref{UpLimmore}, where we consider the background fluctuations by generating isotropic distributions of events in the sky and compute each time $N'_{\rm on}=N_{\rm on} + N_{\rm s}$ and $N_{\rm off}$ (with $N_{\rm s}$, the generated source events). Fig. \ref{UpLimmore} shows the computed event rate upper limit for the Bayesian and Feldman-Cousins methods as a function of the generated $N_{\rm s}$ events. We see that the decrease of the slopes of the upper limit determinations (Fig. \ref{ULcomp}) result in an upper limit that can fall below the actual generated rate. This problem occurs less often for the Bayesian method because the decrease of the slope is less steep compared to Feldman-Cousins. Moreover, as we assume an uniform prior, the Bayesian limit is equal to the Feldman-Cousins for large over-fluctuations and this translates to Fig. \ref{UpLimmore} by having the same values for the largest upper limits of the rate.

Note that the obtained rate does not take into account the reconstruction efficiency. The latter is taken into account by converting the source rate upper limit into a flux upper limit by means of the so called Effective Area, $\mathrm{A_{eff}}$, which is defined as
\be
\mathrm{A_{eff}}=\frac{\mathrm{observed\ event\ rate}}{\mathrm{incoming\ flux}}.\nonumber
\label{Aeffective}
\ee

For the current analysis we use the angle averaged Effective Area determined from a simulated $E^{-2}$ spectrum \cite{diffuse}, taking into account the observed energy estimate for each individual observed event \cite{data}. The median value corresponds to A$_\mathrm{eff} = 2.2 \times 10^6\,$cm$^2$ over the considered energy range. Our analysis is performed on a circular area of 4$^\circ$ centered on each of the 10 sources, representing in total $10 \times 0.0153\ {\rm sr} = 0.153\ {\rm sr} $. From the result of $s_{\rm u. l.}$ and taking the effective area and the size of the on-source region into account we arrive at a 90\% upper limit for an $E^{-2}$ signal flux of $\Phi_\mathrm{u.l.} = \dfrac{s_\mathrm{u.l.}}{\mathrm{A_{eff}}\cdot 0.153} = 3.6 \times 10^{-12}\,\rm{TeV}\,\rm{s}^{-1}\,\rm{cm}^{-2}\,\rm{sr}^{-1}$.

However, this flux upper limit does not take into account the effect of neutrino oscillations. At the source, astrophysical models predict a flavor ratio of $\nu_\mu : \nu_e : \nu_\tau = 2:1:0$. Assuming maximum oscillation we expect to observe at Earth $\nu_\mu : \nu_e : \nu_\tau = 1:1:1$. A tiny fraction of the $\nu_\tau$ will produce a muon which might also be detected in IceCube and as such have entered our event sample. However, we will neglect this effect since it is marginal and would require a special simulation which is beyond the scope of this paper. So our final value for the 90\% upper limit for a E$^{-2}$ signal flux is:
\be
\Phi_\mathrm{u.l.}= 7.2\times 10^{-12}\,\rm{TeV}\,\textrm{s}^{-1}\,\textrm{cm}^{-2}\,\textrm{sr}^{-1}.\nonumber
\ee

For consistency checking, we can compare our limit to the result published by the IceCube Collaboration \cite{diffuse}, which was obtained with a different analysis concerning a search for a diffuse high-energy neutrino flux in the full Northern hemisphere: $\Phi_\mathrm{u.l.}= 8.9\times10^{-12}\,\rm{TeV}$ s$^{-1}\,$cm$^{-2}\,$sr$^{-1}$, which is comparable to our result. In that analysis the same data set was used and since the ten blazars we studied are randomly located in the sky and have not been selected based on any (astro)physical characteristics, the $4^{\circ}$ windows around them represent a fair sample of the sky which can be used to compare to a diffuse search.

\section{Conclusion and Outlook}

In this paper we have discussed a statistical method to analyse point sources using data from a neutrino telescope following Bayesian inference. Using the observable $\psi$, we have indicated how to assess the significance of a possible signal in the data by comparing it to the $\psi$ distribution expected for an isotropic background. We have shown how to obtain upper limits for the corresponding flux in case the observation does not lead to a significant signal detection. Our calculations are similar to \cite{loredo,greg} but we have made no approximations in the final results and thus this method can be applied to low counting observations.

Applying this method we have analysed the public IceCube 40-string configuration data for 10 nearby blazars located in the Northern sky. From our analysis it was also seen that the on-source data is consistent with an isotropic background only hypothesis. Therefore we have determined a 90\% upper limit, which has been compared to the upper limits obtained using the same data set but applying the Feldman-Cousins and the Rolke \textit{et al.} methods. Simulating a signal from a source, by artificially changing the number of events in the on-source region, we have shown that the Bayesian limits are similar to the Rolke \textit{et al.} calculations for small difference in the number of events between the on-source and off-source region and tend to the Feldman-Cousins limits as this difference in the number of events increases. We have shown that in the case of under-fluctuations in the background the Bayesian method is better protected.

It is our intention to extend the current method also for non-steady sources like for instance GRBs. Apart from providing a signal significance for discovery \cite{NvE}, this should also provide a mechanism to accurately determine flux upper limits. For flaring sources we do not know the time window in which the neutrinos are emitted. If we take a time window large enough to cover all possible scenarios for neutrino emission, we would obtain a rate which is not the actual one (because of the existence of time intervals with and without neutrino emission within our time window). The proper extension of the method is currently under study.

\section*{Acknowledgements}

The authors would like to thank the IceCube Collaboration for providing the public data used in this report to evaluate our analysis method. This research was performed with financial support from the Odysseus programme of the Flemish Foundation for Scientific Research (FWO) under contract number G.0917.09.

\newpage
\section*{Appendix A1}

The normalisation factor of the posterior background rate PDF given in Eq. \rf{b13} may be written as:
\begin{equation}
p(N_\textrm{off}|I)=\frac{1}{b_\textrm{max}\,N_\textrm{off}!}\int_0^{b_\textrm{max}}(bT_\textrm{off})^{N_\textrm{off}}\,e^{-bT_\textrm{off}}\,\textrm{d}b.
\label{bckgd}
\end{equation}
The integral part can be expressed as the so-called Incomplete Gamma function given by:
\begin{equation}
\gamma(a,x)=\int_0^xe^{-t}\,t^{a-1}\textrm{d}t.
\label{incompletegamma}
\end{equation}
Using this expression, we can rewrite Eq. \rf{bckgd} as follows:
\begin{eqnarray}
p(N_\textrm{off}|I)&=&\frac{1}{b_\textrm{max}\,N_\textrm{off}!}\int_0^{b_\textrm{max}}T_\textrm{off}(bT_\textrm{off})^{N_\textrm{off}}\,e^{-bT_\textrm{off}}\,\frac{\textrm{d}(bT_\textrm{off})}{T_\textrm{off}}\\
&=& \frac{1}{b_\textrm{max}}\frac{\gamma(N_\textrm{off}+1,b_\textrm{max}T_\textrm{off})}{N_\textrm{off}!\,T_\textrm{off}},
\end{eqnarray}
which is the expression reflected in Eq. \rf{b14}.
\section*{Appendix A2}
According to Eq. \rf{b19} the posterior PDF for the source rate alone is given by:
\begin{equation}
p(s|N_\textrm{on},I)=\int_0^{b_\textrm{max}} p(s,b|N_\textrm{on},I)\,\textrm{d}b,
\label{postPDFsrate}
\end{equation}
where
\begin{equation*}
p(s,b|N_\textrm{on},I)=\frac{p(s | b,I) p(b | I) p(N_\textrm{on} | s,b,I)}{p(N_\textrm{on} | I)}.
\end{equation*}
Substitution of the various expressions given in Section 4 yields:
\be
p(s,b|N_\textrm{on},I)=\frac{\frac{1}{s_\textrm{max}}\cdot\frac{T_\textrm{off}(bT_\textrm{off})^{N_\textrm{off}}e^{-bT_\textrm{off}}}{\gamma(N_\textrm{off}+1, b_\textrm{max}T_\textrm{off})} \cdot \frac{(b+s)^{N_\textrm{on}}T_\textrm{on}^{N_\textrm{on}}e^{-(b+s)T_\textrm{on}}}{N_\textrm{on}!} }{\int_0^{b_\textrm{max}}\int_0^{s_\textrm{max}}\frac{1}{s_\textrm{max}}\cdot\frac{T_\textrm{off}(bT_\textrm{off})^{N_\textrm{off}}e^{-bT_\textrm{off}}}{\gamma(N_\textrm{off}+1, b_\textrm{max}T_\textrm{off})}\cdot \frac{(b+s)^{N_\textrm{on}}T_\textrm{on}^{N_\textrm{on}}e^{-(b+s)T_\textrm{on}}}{N_\textrm{on}!} \,\textrm{d}b\,\textrm{d}s}.
\label{annexe2}
\ee
Combination of Eqs. \rf{postPDFsrate} and \rf{annexe2} yields:
\begin{eqnarray}
p(s|N_\textrm{on},I)&=&\int_0^{b_\textrm{max}}\frac{\frac{1}{s_\textrm{max}}\cdot \frac{T_\textrm{off}(bT_\textrm{off})^{N_\textrm{off}}e^{-bT_\textrm{off}}}{\gamma(N_\textrm{off}+1, b_\textrm{max}T_\textrm{off})}\cdot \frac{(b+s)^{N_\textrm{on}}T_\textrm{on}^{N_\textrm{on}}e^{-(b+s)T_\textrm{on}}}{N_\textrm{on}!} }{\int_0^{b_\textrm{max}}\int_0^{s_\textrm{max}}\frac{1}{s_\textrm{max}}\cdot \frac{T_\textrm{off}(bT_\textrm{off})^{N_\textrm{off}}e^{-bT_\textrm{off}}}{\gamma(N_\textrm{off}+1, b_\textrm{max}T_\textrm{off})}\cdot \frac{(b+s)^{N_\textrm{on}}T_\textrm{on}^{N_\textrm{on}}e^{-(b+s)T_\textrm{on}}}{N_\textrm{on}!} \,\textrm{d}b\,\textrm{d}s}\,\textrm{d}b\nonumber\\
&=&\frac{\int_0^{b_\textrm{max}}\frac{1}{s_\textrm{max}}\cdot \frac{T_\textrm{off}(bT_\textrm{off})^{N_\textrm{off}}e^{-bT_\textrm{off}}}{\gamma(N_\textrm{off}+1, b_\textrm{max}T_\textrm{off})}\cdot \frac{(b+s)^{N_\textrm{on}}T_\textrm{on}^{N_\textrm{on}}e^{-(b+s)T_\textrm{on}}}{N_\textrm{on}!} \,\textrm{d}b}{\int_0^{b_\textrm{max}}\int_0^{s_\textrm{max}}\frac{1}{s_\textrm{max}}\cdot \frac{T_\textrm{off}(bT_\textrm{off})^{N_\textrm{off}}e^{-bT_\textrm{off}}}{\gamma(N_\textrm{off}+1, b_\textrm{max}T_\textrm{off})}\cdot \frac{(b+s)^{N_\textrm{on}}T_\textrm{on}^{N_\textrm{on}}e^{-(b+s)T_\textrm{on}}}{N_\textrm{on}!} \,\textrm{d}b\,\textrm{d}s}\label{annexe2bis}\\
&\equiv&\frac{\textrm{A}}{\textrm{B}}\nonumber.
\end{eqnarray}
Considering the numerator of Eq. \rf{annexe2bis}, we obtain:
\begin{eqnarray*}
\textrm{A}&=&\int_0^{b_\textrm{max}}\frac{1}{s_\textrm{max}}\cdot \frac{T_\textrm{off}(bT_\textrm{off})^{N_\textrm{off}}e^{-bT_\textrm{off}}}{\gamma(N_\textrm{off}+1, b_\textrm{max}T_\textrm{off})}\cdot \frac{(b+s)^{N_\textrm{on}}T_\textrm{on}^{N_\textrm{on}}e^{-(b+s)T_\textrm{on}}}{N_\textrm{on}!} \,\textrm{d}b\\
&=&\frac{1}{s_\textrm{max}}\,\frac{T_\textrm{on}^{N_\textrm{on}}T_\textrm{off}^{N_\textrm{off}+1}}{N_\textrm{on}!}\frac{1}{\gamma(N_\textrm{off}+1, b_\textrm{max}T_\textrm{off})}\int_0^{b_\textrm{max}} (b+s)^{N_\textrm{on}}b^{N_\textrm{off}}e^{-(b+s)T_\textrm{on}}e^{-bT_\textrm{off}}\,\textrm{d}b.
\end{eqnarray*}
Using the Newtonian Binomial, $(a+b)^n=\sum_{i=0}^{n}\frac{n!}{i!(n-i)!}a^{n-i}b^i$, we find:
\begin{equation*}
\textrm{A}=\frac{1}{s_\textrm{max}}\,\frac{T_\textrm{on}^{N_\textrm{on}}T_\textrm{off}^{N_\textrm{off}+1}}{N_\textrm{on}!}\frac{1}{\gamma(N_\textrm{off}+1, b_\textrm{max}T_\textrm{off})}\sum_{i=0}^{N_\textrm{on}}\frac{N_\textrm{on}!\,s^i\,e^{-sT_\textrm{on}}}{i!(N_\textrm{on}-i)!}\int_0^{b_\textrm{max}}b^{N_\textrm{on}+N_\textrm{off}-i}e^{-b(T_\textrm{on}+T_\textrm{off})}\,\textrm{d}b.
\end{equation*}
Using the Incomplete Gamma function, Eq. \rf{incompletegamma}, and simplifying the above equation, we finally find
\begin{equation}
\textrm{A}=\frac{1}{s_\textrm{max}}\,\frac{T_\textrm{on}^{N_\textrm{on}}T_\textrm{off}^{N_\textrm{off}+1}}{\gamma(N_\textrm{off}+1, b_\textrm{max}T_\textrm{off})}\sum_{i=0}^{N_\textrm{on}}\frac{s^i\,e^{-sT_\textrm{on}}}{i!(N_\textrm{on}-i)!}\,\frac{\gamma(N-i+1, u_\textrm{max})}{(T_\textrm{on}+T_\textrm{off})^{N_\textrm{on}+N_\textrm{off}-i+1}},
\end{equation}
where $u_\textrm{max}=b_\textrm{max}(T_\textrm{on}+T_\textrm{off})$ and $N=N_\textrm{on}+N_\textrm{off}$.\\
Applying the same procedure, we obtain for the denominator:
\begin{eqnarray*}
\textrm{B}&=&\int_0^{b_\textrm{max}}\int_0^{s_\textrm{max}}\frac{1}{s_\textrm{max}}\cdot \frac{T_\textrm{off}(bT_\textrm{off})^{N_\textrm{off}}e^{-bT_\textrm{off}}}{\gamma(N_\textrm{off}+1, b_\textrm{max}T_\textrm{off})}\cdot \frac{(b+s)^{N_\textrm{on}}T_\textrm{on}^{N_\textrm{on}}e^{-(b+s)T_\textrm{on}}}{N_\textrm{on}!} \,\textrm{d}b\,\textrm{d}s\\
&=&\int_0^{s_\textrm{max}}A\,\textrm{d}s\\
&=&\frac{1}{s_\textrm{max}}\,\frac{T_\textrm{on}^{N_\textrm{on}}T_\textrm{off}^{N_\textrm{off}+1}}{\gamma(N_\textrm{off}+1, b_\textrm{max}T_\textrm{off})}\sum_{j=0}^{N_\textrm{on}}\frac{1}{j!(N_\textrm{on}-j)!}\,\frac{\gamma(N-j+1, u_\textrm{max})}{(T_\textrm{on}+T_\textrm{off})^{N_\textrm{on}+N_\textrm{off}-j+1}}\int_0^{s_\textrm{max}}s^j\,e^{-sT_\textrm{on}}\,\textrm{d}s\\
&=&\frac{1}{s_\textrm{max}}\,\frac{T_\textrm{on}^{N_\textrm{on}}T_\textrm{off}^{N_\textrm{off}+1}}{\gamma(N_\textrm{off}+1, b_\textrm{max}T_\textrm{off})}\sum_{j=0}^{N_\textrm{on}}\frac{1}{j!(N_\textrm{on}-j)!}\,\frac{\gamma(N-j+1, u_\textrm{max})}{(T_\textrm{on}+T_\textrm{off})^{N_\textrm{on}+N_\textrm{off}-j+1}}\frac{\gamma(j+1,s_\textrm{max}T_\textrm{on})}{T_\textrm{on}^{j+1}}.
\end{eqnarray*}
Which finally gives, after simplification, for the posterior source rate probability density function:
\begin{equation}
p(s|N_\textrm{on},I)=\frac{e^{-sT_\textrm{on}}\,\sum_{i=0}^{N_\textrm{on}}\frac{s^i\,(T_\textrm{on}+T_\textrm{off})^i\,\gamma(N-i+1, u_\textrm{ max})}{i!(N_\textrm{on}-i)!}}{\sum_{j=0}^{N_\textrm{on}}\frac{(T_\textrm{on}+T_\textrm{off})^j\,\gamma(j+1,s_\textrm{ max}T_\textrm{on})\,\gamma(N-j+1, u_\textrm{ max})}{j!(N_\textrm{on}-j)!\,T_\textrm{on}^{j+1}}}.
\end{equation}

\end{document}